\documentclass{sig-alternate}
\usepackage{graphicx}
\usepackage{epstopdf}
\usepackage{etoolbox}
\usepackage{booktabs}
\makeatletter
\patchcmd{\maketitle}{\@copyrightspace}{}{}{}
\makeatother
\usepackage[english]{babel}

\begin{document}

\title{Plagiarism Detection - State-of-the-art systems (2016) and evaluation methods}

\numberofauthors{1} 
%
\author{
%
%
\alignauthor
Christina Kraus\\
       \affaddr{Technische Universit\"at Berlin}\\
       \affaddr{Database Systems and Information Management Group}\\
       \email{c.kraus@campus.tu-berlin.de}
}

\maketitle
\begin{abstract}
Plagiarism detection systems comprise various approaches that aim to create a fair environment for academic publications and appropriately acknowledge the authors' works. 
While the need for a reliable and performant plagiarism detection system increases with an increasing amount of publications, current systems still have shortcomings. Particularly intelligent research plagiarism detection still leaves room for improvement. An important factor for progress in research is a suitable evaluation framework.
In this paper, we give an overview on the evaluation of plagiarism detection. We then use a taxonomy provided in former research, to classify recent approaches of plagiarism detection. Based on this, we asses the current research situation in the field of plagiarism detection and derive further research questions and approaches to be tackled in the future.
\end{abstract}

\section{Status of 2012}
The work of Alzahrani et al.\cite{5766764} structures former research to present an overview on the current research status in plagiarism detection. It differentiates types of plagiarism, plagiarism detection system frameworks, textual features that are needed to apply plagiarism detection methods and finally different plagiarism detection methods. 
In a first step they  present a new taxonomy of plagiarism in which they identify two main types, literal and intelligent plagiarism. They consider the latter as a ``serious academic dishonesty'' as well as more common in the academic environment. Both types further divide into: Exact, near and modified copies in the case of literal plagiarism and text manipulation, translation and idea adoption for intelligent plagiarism. 

\begin{figure}
	\vspace{2ex}
	\includegraphics[width=0.9\linewidth]{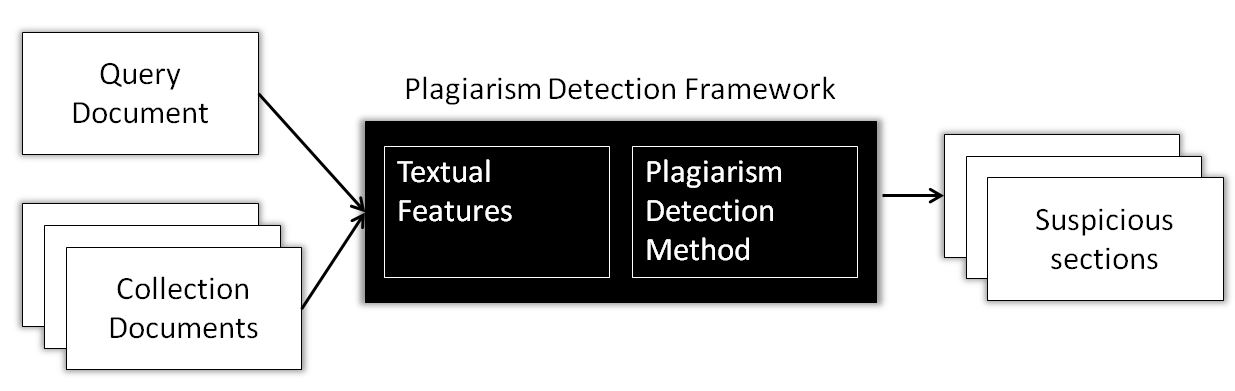}
	\label{pic}
	\caption{Extrinsic Plagiarism Detection Framework}
\end{figure}

Plagiarism detection frameworks describe different approaches for the general functionality of a plagiarism detection system and could be described as different tasks in plagiarism detection that can be applied separately or combined. 
There are three options of detection frameworks: Extrinsic, Intrinsic and Crosslingual plagiarism detection. Extrinsic plagiarism detection takes a source document collection into consideration. The task of the plagiarism detector is to find supicious sections in a query document in relation to that source document collection. This framework includes the task of finding relevant source documents. Figure \ref{pic} shows a model of an extrinsic plagiarism detection framework.Intrinsic plagiarism detection aims to detect supsicious sections of the query document without any additional source, for example by changes in writing style within the same document. Lastly, crosslingual frameworks attempt to detect plagiarism across different languages.

In order to identify potentially plagiarized sections of a document, plagiarism detection frameworks use various textual features to characterize a text and thereby enable comparability to other texts. Lexical features can be N-grams or Word-grams. Syntactical features are bigger portions of texts, such as chunks, sentences or POS. Semantic features exceed the actual content of the text and take into consideration abstract concepts rather than concrete words. Synonyms and antonyms are examples. The existance of structural features depends on the type of text. Academic texts follow a structure based on their sections and paragraphs. Another group of features, stylometric features, emerges from statistics on the previously mentioned groups, such as frequency of words, average sentence length, synonyms, average paragraph length.  

Comparing these features for different documents or text sections requires different methods, which are then suitable to detect different types of plagiarism. Character-based, vector-based and syntax-based methods are suitable for literal plagiarism detection. Similarly, structural-based combined with vector-based methods. Semantic and fuzzy methods have been used to identify intelligent plagiarism up to now and stylometric methods for the intrinsic detection of plagiarism. As we use this paper as a starting point for the categorization and evaluation of new detection approaches, we refer to it as ``the base paper'' in the remainder of this paper.

While the structure the paper provides is clear and covers most approaches, it neglects some aspects that are important for future research in this field: First of all, performance in terms of detection quality is not evaluated. Secondly, computational cost is an important issue, especially when it comes to extrinsic plagiarism detection, as this brings the need to compare the paper in question to every other document in the source collection. As this source collection ideally includes all related research, it can be very large. 
The differentiation of all these methods leads to the conclusion that different types of plagiarsm require different detection methods. Another question that arises is, how to identify different plagiarism methods in a real-life scenario, where we do not know what to expect, for example a conference paper submission tool.
We address these issues in the remainder of this paper in the following way: First we present new approaches that evolved after 2012 and integrate them into the given structure (as far as possible). In addition, we evaluate them comparatively regarding their result quality and compuational performance, whenever this information is available. In a next step, we introduce the PAN-PC competition, a yearly held competition for plagiarism detection systems. We compare the result evolution of the last 5 years and present the winner of the competition in 2015. Based on these developments, we analyze the research situation in plagiarism detection and derive weaknesses as well as further research questions to be addressed in the future.

\section {Evaluation of Plagiarism Detection systems}
The most common task in plagiarism detection is extrinsic plagiarism detection. This includes the consideration of a source collection of documents or reference corpus in which the origin of a potentially plagiarized document would be located. In order to assess the development of a research field, we need an evaluation framework that allows to qualitatively compare various approaches over years and moreover represents a fairly good estimation of the real-world-scenario that the research question addresses. Potthast et al.\cite{evaluation} attempt to answer this question with the creation of the PAN-PC-Corpus in 2010. The need for such a framework arises from the analysis of former research and the findings that the used corpora as well as the evaluation measures are diverse and non-consistent over time. They state that a corpus that contains real plagiarism examples, brings ethical and legal issues and therefore favour the artificial creation of a document corpus for evaluation. 
The PAN-PC competition is a yearly held competition on digital forensics. It includes multiple tasks, among them the retrieval of source documents, as well as plagiarism detection and has the aim to create a representative evaluation framework for this research field. Therefore, it provides a document corpus on which participants evaluate their approaches. The corpus contains documents that consist of copy\&pasted parts of books. It includes two types of plagiarism: artificial plagiarism and simulated plagiarism. Artificial plagiarism describes the reproduction of document parts in other documents (exact copies), while simulated plagiarism cases reproduce text through manual paraphrasing. In addition the corpus contains cases of manually and automatically translated text reproduction. 
A common way to evaluate the quality of plagiarism detection systems are the measures of recall and precision. The F-Measure is the harmonic mean between recall and precision and therefore caputures the detection quality in a single value. To capture the concept of granularity the PAN-PC competition introduces the PlagDet measure, defined as:
\begin{align*}
\text{plagdet}(S,R) = \frac{F_{1}}{\log_{2}(1+\text{gran}(S,R))}\end{align*}
Here, gran refers to the so called granularity of the detector that describes the fact that some approaches detect a single occurence of plagiarism as multiple or overlapping occurences. 

\section{New approaches of Plagiarism Detection}
As the groups and detection frameworks in the paper are disjunct from concrete approaches, it is mostly possible to classify new approaches in the same manner. Following
we introduce 6 new approaches to plagiarism detection. 

\subsection{Citation-based plagiarism detection}
In 2011 Gipp et al. \cite{gipp2011comparative} introduce a new approach to the problem of plagiarism detection. By 2014, a more elaborate version of this approach has been published \cite{gipp2014citation}. Citation-based plagiarism detection characterizes a document by its citation sequence, rather than the text itself. Based on a sequential pattern analysis the authors compare different algorithms, such as the longest common citation sequence of two documents or citation chunking. This method comes with two important advantages: The approach is language-independant and, by focussing on the reference list, it drastically reduces complexity compared to the comparison of full documents. Gipp et al. evaluate their approach on the PubMed Central Open Access Subset corpus. This corpus, limited to medical publications, represents a real-life scenario, as real plagiarism cases are searched for and not imitated. The source collection comprises 185,170 documents. To evaluate the approach the authors conduct a user study, where individuals in different positions (undergraduate, graduate, experts) assign plagiarism scores to given documents to put them in a rank. These ranks are then compared to a plagiarism rank that results from the different algorithms. While this evaluation approach seems reasonable in order to develop a person-like plagiarism detection system it does not allow for a comparison to other approaches. However, the study includes a comparison to character-based approaches and finds that all algorithms in citation-based plagiarism detection perform better than character-based plagiarism detection in the cases of structural and idea plagiarism as well as paraphrasing, but not for copy\&paste and copy\&shake plagiarism
\subsection{Semantic Similarity}
Hussain\&Suryani \cite{Hussain} use an algorithm, known as the $\chi$-Sim algorithm in their approach for plagiarism detection. It derives similarity values between texts based on the statistical evaluation of word co-occurences. As n:n comparisons between documents are costly, the authors reduce time complexity by reducing the amount of documents to compare through a prior title comparison. This complexity reduction method assigns a document to a certain category, so that a new document is only compared to documents in the same category rather than the whole document corpus. For the evaluation, the authors artificially insert plagiarism in the source document collection by rewriting randomly chosen documents from different categories. Thereby, they differentiate in the percentage of plagiarism inserted into a document (varying between 0\%-100\%), which imitates the difference between copy \& paste plagiarism and paraphrasing. For the quality evaluation they compare the approach to Turnitin, a commercial plagiarism detection system using a plagiarism index as a measure. The index indicates the estimated amount of plagiarism for each suspicious document. A comparison to precision \& recall measures is not possible. The experiments result in the fact that Turnitin seems to underestimate plagiarism, especially in documents with a high percentage of plagiarism. While these results sound promising the experimental setting is rather unrealistic due to manually inserted plagiarism and not comparable to other approaches due to the use of a different measure.
\subsection{Semantic Role Labeling}
Semantic role labeling is an approach that calculates the similarity between sentences. The idea is to split sentences and assign their parts to different label groups. A second step consists of the semantic annotation of terms with the help of the semantic lexicon WordNet .This enables to see parallels in sentences such as: ``John eats food.'' and ``Mary drinks water.'' The authors evaluate their approach on a subset of 100 documents of the PAN-PC 09 data set and receive an improvement of more than 35\% for both, precision and recall, compared to the approaches that originally took part in the competition. In a comparison to 5 other plagiarism techniques including semantic, fuzzy and vector-based methods, this approach performs best. In a complexity analysis the approach belongs to the time complexity class of $O(n^2)$ \cite{Osman20121493}. However, this analysis does not include a time estimation for a single document analysis.
\subsection{Latent semantic indexing (LSI)}
Another unmentioned approach in the group of semantic approaches uses the concept of latent semantic indexing in combination with intrinsic plagiarism detection. The former builds a probabilistic model for words based on a SVD-reduced term-document matrix. The evaluation of this approach is restricted to the analysis of the effect of the dimensionality reduction through SVD. More specifically, the analysis concentrates on a performance comparison of two stylometric features, namely ``Average Sentence Length'' and the ``Honore Function'', with a varying number of dimensions in the LSI step. This results in a preferable number of dimensions that lies between 20\% and 40\% \cite{7041107}. There is no information on the character of the applied quality measurement and therefore we cannot draw parallels to other methods.
\subsection{Crosslingual Semantic Approach}
The crosslingual semantic approach uses a semantically annotated graph model for crosslingual plagiarism detection. It establishes semantic relations through the use of WordNet, BabelNet - a multilingual encyclopedic dictionary \cite{babelnet}  and Wikipedia. The resulting knowledge graph consists of the concepts from the document and its neighbour concepts. Each graph characterizes a document and similarity between graphs represents similarity between documents. The authors choose a subset of the PAN-PC 11 corpus for evaluation and the plagdet measure. They receive the best results for English-Spanish translations. With a plagdet value of 0.594 their approach performs better than competition participants of 2011 (same corpus), but worse than competition participants of the next year (new corpus) \cite{multilingualSemantic}.
\subsection{Regular Cross Lingual Approach}
As presented in the origin paper, this approach combines lexical and syntactic features such as character-n-grams without taking semantic annotations into consideration. It further follows the idea to translate a document before applying a monolingual detection approach. For the evaluation the authors again use a subset of the PAN-PC-11 data set. However, the results cannot compete with the ones of the semantic crosslingual approach described earlier in this section \cite{crosslingual}. Especially the recall value of 0.25 represents a poor result compared to all other approaches.

\section {Assessment of the current status}
\subsection {Detection Quality Development}
Regarding the development of the results in the PAN competition, we observe an increase of more than 50\% from 2011 to 2015. Table \ref {tab:table} shows the results of the bestperforming plagiarism detectors for each year since 2011.
\begin{table}
	\center 
	\begin{tabular}{l l}
		
		\toprule
		\textbf{Competition} & \textbf{PlagDet}\\
		\midrule
		3rd PAN PC (2011) & 0.56  \\
		4th PAN PC (2012) & 0.738 \\
		5th PAN PC (2013) & 0.832\\
		6th PAN PC (2014) & 0.878\\
		\bottomrule
	\end{tabular}
	\caption{PAN-PC Competition Result Evolution \cite{stein2011overview, panpc4, panpc5, panpc6}}
	\label{tab:table}
\end{table} The winner of the last PAN competition in 2015 used a tf-idf vector-based approached with a vector space model. The approach also incorporates the type of plagiarism, by analyzing the length of a supsicious section with regard to summarized plagiarism. The parameter tuning depends on this classification. This behavior leads to the assumption that the competitors highly adapt their approaches to the given corpus characteristics. Moreover, many participants compete with the same approach over several years with varying parameters, which strengthens the assumption of a significant parameter fitting process. While this process does not generally negatively influence the results, it might rely too much on the characteristics of the artificially created document corpus. 
\subsection {Research Environment}
While there is no doubt in the need of an evaluation framework for plagiarism detection, our assessment shows that even with the PAN-PC corpus the problems of various measures and resulting incomparability of different approaches still holds. This may be a result of the following issues in the PAN-PC corpora: Gipp et al. criticize that the proportion of manually paraphrased towards artificially plagiarized text is insufficient, when compared to real-life situations. This leads to a discrepancy in the performance of plagiarism detection approaches regarding literal vs. intelligent plagiarism. Another expression of critique on the corpus refers to the random insertion of plagiarism, meaning that documents that contain reproduced sections do not necessarily belong to the same topic. This is inprobable in a real-life scenario \cite{Hussain}.
\subsection{Future Work}
Even though the topic of plagiarism detection has been around since the 70's, we still observe a lot of effort put into the development of new detection methods and new evolving ideas. We found that despite the fact that an extensive evaluation platform for plagiarism detection research has been established through the PAN-PC competition there are parallel developments in this research field. While competition participants put a lot of effort in tuning their systems for the competition corpus, new approaches often use different corpora as well as measures and thereby impede direct comparisons to former systems or competition participants.  Different steps could be taken to improve the developments in this research field. First of all, evaluation measures should be adjusted to be similar. Some researchers mention the unsuitability of the PAN-PC corpora for their approach. Joint work on corpus creation, could help to improve these and help to approximate real-life scenarios. Furthermore, desirable improvements in this research field include the consideration of performance, as this is essential for a real-life-scenrio with a reference corpus of reasonable size.
\subsection{Acknowledgements}
This paper was written in the context of the Database Seminar WS 15/16 at the Database and Information Management Group at Technische Universit\"at Berlin and
supervised by Moritz Schubotz.

\bibliographystyle{abbrv}
\bibliography{sigproc}  

\end{document}